\documentclass[a4paper,12pt]{article}
\usepackage{bm}
\usepackage{amsmath,amsfonts,amssymb}
\newcommand{\ve}{\bm{\xi}}
\newcommand{\vp}{\bm{\xi'}}
\newcommand{\vs}{\bm{\xi}_{*}}
\newcommand{\vsp}{\bm{\xi'}_{*}}
\newcommand{\n}{\mathbf{n}}
\newcommand{\bV}{\mathbf{V}}
\newcommand{\bx}{\mathbf{x}}
\newcommand{\bC}{\mathbf{C}}
\newcommand{\cE}{\mathcal{E}}
\newcommand{\iiW}{S}
\newcommand{\rea}{\mathbb{R}}
\newcommand{\irn}[1]{\int_{\rea^{#1}}}
\newcommand{\p}{\, .}
\newcommand{\sv}{\, ,}
\newcommand{\eq}[1]{(\ref{#1})}
\newcommand{\dm}{\displaystyle}
\newcommand{\oot}{\mbox{${\frac{1}{2}}$}}
\author{Armando Majorana
\\[5pt] Dipartimento di Matematica e Informatica
\\ Viale A. Doria 6, 95125 Catania, Italy}
\title{}
\title{A deterministic numerical model for the nonlinear Boltzmann equation}
\begin{document}
\maketitle
%
%
\begin{abstract}
We propose a new deterministic numerical scheme, based on the discontinuous Galerkin
method, for solving the Boltzamnn equation for rarefied gases.
The new scheme guarantees the conservation of the mass, momentum and energy.
We avoid any stochastic procedures in the treatment of the collision operator of the
Boltzmamn equation.
\end{abstract}
MSC-class: 76P, 82C40 (Primary) 65M60 (Secondary)
%
%
%
\section{Introduction}
The classical Boltzmann kinetic equation describes neutral particle transport phenomena.
Today numerical solutions of the Boltzmann equation are requested to solve problems
in different fields of real-world applications.
There are two classes of computational methods, which are used to solve the
kinetic equation. 
In the first of techniques, the well-known Direct Simulation Monte Carlo (DSMC) method,
the molecular collisions are considered on a probabilistic rather than a deterministic
basis.
The literature of the applications of this method is very vast.
In the second class, deterministic methods, the Boltzmann equation is
discretized using a variety of methods and then solved directly or iteratively. 
The computational complexity is high due to the large number of independent variables.
This heavy computational cost explains why kinetic equations are
traditionally simulated by the Direct Simulation Monte Carlo methods.
As examples of papers dealing with deterministic schemes for the Boltzmann equation, we
cite some references \cite{BH}, \cite{SOA}, \cite{RS} and the book \cite{Aristov}, but
there are many other interesting works.

In recent years, deterministic solvers to the system, given by Boltzmann equation coupled
to Poisson equation, describing of electron flow in semiconductors were considered in the
literature (see, for example, \cite{carr03}, \cite{cgms06}, \cite{Cheng_08}).
These methods provide accurate results which, in general, agree well with those obtained
from DSMC simulations, sometimes at a comparable or even less computational time. 
The discontinuous Galerkin (DG) method, which is a finite element method using
discontinuous piecewise polynomials as basis functions and numerical fluxes based on
upwinding for stability, seems to be good for solving also kinetic equations.   
The method has the advantage of flexibility for arbitrary unstructured meshes, with a
compact stencil, and with the ability to easily accommodate arbitrary {\it hp-}adaptivity.
For more details about DG scheme for convection dominated problems, we refer to the review
paper \cite{dgsurvey}. 

The starting point of this paper is a weak formulation of the Boltzmann equation
\cite{C88}, \cite{C90}
$$
 \frac{\partial f}{\partial t} + \ve \cdot \frac{\partial f}{\partial \bx} = Q(f,f) \p
$$
If we multiply both sides of the equation by a test function $\phi(\bx, \ve)$ and we
integrate with respect to the coordinates $\bx$ and  the velocity $\ve$, 
then we obtain the equation
\begin{eqnarray*}
&&
 \frac{\partial \mbox{ }}{\partial t} \irn{3} \int_{X}
 f(t,\bx, \ve) \, \phi(\bx, \ve) \: d \bx \, d \ve +
 \irn{3} \int_{X} \ve \cdot 
 \frac{\partial f}{\partial \bx}(t,\bx, \ve) \, \phi(\bx, \ve) \: d \bx \, d \ve 
 \\
 && \mbox{} =
\irn{3} \int_{X} Q(f,f)(t,\bx, \ve) \, \phi(\bx, \ve) \: d \bx \, d \ve \, ,
\end{eqnarray*}
where $X$ denotes the $\bx$-domain.
It is necessary an integration by parts to move the derivative with respect to $\bx$ from
the function $f$ to $\phi$. This requires some information on the domain $X$ and the test
function. We do not describe this step, because, in this paper, we will study only the
r.h.s of the Boltzmann equation.
The test function $\phi$ belongs to a suitable chosen finite dimensional space, which is
also used to find an approximation of the unknown $f$.

The plan of the paper is the following.
In Section 2, we will introduce the Boltzmann equation. Section 2, 3 and 4 will be
devoted to the weak form of the collision operator and a new modified version.
Section 5 will show a simple approximate distribution function $f$ to be used in the
framework of the discontinuous Galerkin method.
Conclusions and future work are given in Section 6.
%
%
\section{The Boltzmann equation}
The purpose of this section is to briefly introduce the classical nonlinear Boltzmann
equation for monatomic gases, to recall well-known properties and to derive simple
results.
According to the standard notation, the Boltzmamn equation takes the form
\begin{equation}
\frac{\partial f}{\partial t} + \ve \cdot \frac{\partial f}{\partial \bx} =
\irn{3} \irn{3} \irn{3}  W(\ve, \vs | \vp, \vsp) 
\left( f' f'_{*} - f f_{*} \right)  d \vs \, d \vp \, d \vsp  \p \label{eqb}
\end{equation}
The one-particle distribution function $f$ depends on time $t$, position $\bm{x}$ and
velocity $\ve$.
The kernel $W$ of the collision operator is defined by
\begin{equation}
W(\ve, \vs | \vp, \vsp) = K(\n \cdot \bV, |\bV|) \,
\delta(\ve + \vs - \vp - \vsp) \,
\delta(|\ve|^{2} + |\vs|^{2} - |\vp|^{2} - |\vsp|^{2})
 \label{Wv4}  
\end{equation}
where
\begin{equation}
\n = \dfrac{\ve - \vp}{|\ve - \vp|}  \quad \mbox{and} \quad 
\bV = \ve - \vs \p \label{nV}
\end{equation}
The function $K$ is related to the interaction law between colliding particles. 
The Dirac distributions guarantee momentum and energy conservation during the binary
collisions.
It is immediate to verify that the following symmetry properties
$$
W(\ve, \vs | \vp, \vsp) = W(\vp, \vsp | \ve, \vs)  \quad \mbox{and} \quad
W(\ve, \vs | \vp, \vsp) = W(\vs, \ve | \vsp, \vp)
$$
hold.
The collision operator of the Boltzmann equation \eq{eqb} is usually written in different
way, since Dirac distributions are employed and a five-fold integral is derived.
In this paper we find useful the form of collision term given in \eq{eqb} or the
partially reduced integral operator, where only the Dirac distribution describing
momentum conservation is used to reduce the order of integration.

The integration with respect to the variables $\vs$ and $\vsp$ may be performed ab initio
in the lost term of the collision operator, since the unknown $f$ is not involved.
To this aim, we must consider the total cross section
\begin{equation}
 \irn{3} \irn{3} W(\ve, \vs | \vp, \vsp) \: d \vp \, d \vsp \p
\label{intW} 
\end{equation}
It is related to the collision frequency. It is easy to prove (see Appendix A) that
the integral \eq{intW} is a function of $\ve$ and $\vs$ only through $|\bV|$ and it
is equal to
\begin{equation}
\iiW(|\bV|) =  
\frac{\pi}{4} |\bV| \int_{-1}^{1}
K_{\bC}(\oot |\bV|^{2} \mu , |\bV|) \: d \mu \sv
\label{iWmu}
\end{equation}
where
$$
 K_{\bC}(\zeta, |\bV|) = 
 K \! \left(\sqrt{\oot |\bV|^{2} - \zeta} \, , |\bV| \right) .
$$
%
%
%
\subsection{A modified kernel}
The numerical treatment of a kinetic equation by means of finite differences or
elements requires a bounded domain for the velocity space. We introduce a suitable
characteristic function in the kernel of the collision operator, such that there exists a
bounded domain $\Omega$ so that if $f(0,\bx,\ve) = 0$ for every $\ve \not \in \Omega$ and
$\bx \in X$, then $f(t,\bx,\ve) = 0$ for every $\ve \not \in \Omega$, $\bx \in X$  and for
all time $t$.
Let $\cE$ be a positive real number.
We define the function $ \chi_{\cE} : \rea^{3} \times \rea^{3} \rightarrow \rea$ as
follows
$$
 \chi_{\cE}(\ve, \vs) = 
\left\lbrace 
\begin{array}{ll}
1 &  \mbox{if } |\ve|^{2} + |\vs|^{2} \leq \cE \\[5pt]
0 & \mbox{otherwise}
\end{array}
\right. \sv
$$
and
\begin{equation}
 W_{\cE}(\ve, \vs | \vp, \vsp) = \chi_{\cE}(\ve, \vs) \, W(\ve, \vs | \vp, \vsp)
 \p
\end{equation}
It is immediate to see that
\begin{equation}
\irn{3} \irn{3} W_{\cE}(\ve, \vs | \vp, \vsp) \: d \vp \, d \vsp =
 \chi_{\cE}(\ve, \vs) \, \iiW(|\bV|) \p
\end{equation}
It is evident that the new modified kernel $W_{\cE}$ satisfies the same properties of
symmetry of true kernel $W$. 
Moreover, $W_{\cE}$ guarantees that if the particles, before the impact, have velocities
such that $|\ve|^{2} + |\vs|^{2} \leq \cE$, then the velocities after the impact satisfy
the inequality $|\vp|^{2} + |\vsp|^{2} \leq \cE$.
Using this kernel, for instance, we can choose $\Omega = \left\lbrace \ve \in
\rea^{3} \: : |\ve|^{2} \leq \cE \right\rbrace$.
We can use a smooth function instead of the characteristic function $\chi_{\cE}$ to
modify the kernel and we obtain the same previous conclusions, but this is useless in our
numerical approach. 
\subsection{Weak form of the collisional operator}
Let $\phi : \rea^{3}  \rightarrow \rea$ be a measurable function. Assuming the existence
of the integrals, we can recover the well-known result
\begin{eqnarray}
&&
\irn{3} \left[ \irn{3} \irn{3} \irn{3}  W_{\cE}(\ve, \vs | \vp, \vsp) 
\left( f' f'_{*} - f f_{*} \right)  d \vs \, d \vp \, d \vsp \right] 
\phi(\ve) \: d \ve  \label{Qfpsi}
\\[7pt]
&&
\mbox{}  \hspace{-20pt} = \dfrac{1}{2}
\irn{3} \irn{3} \irn{3} \irn{3}  W(\ve, \vs | \vp, \vsp) \, \chi_{\cE}(\ve, \vs)
\left[ \phi' + \phi'_{*} - \phi - \phi_{*} \right] 
f \,  f_{*} \: d \vp \, d \vsp d \vs \, d \ve \p \nonumber
\end{eqnarray}
We define
\begin{equation}
 G(\phi; \ve , \vs) = \irn{3} \irn{3}  W(\ve, \vs | \vp, \vsp)
\left[ \phi(\vp) + \phi(\vsp) \right] d \vp \, d \vsp \p \label{Gphi}
\end{equation}
It is evident that $ G(\phi; \ve , \vs) =  G(\phi; \vs , \ve)$.
We now write the integral \eq{Qfpsi} in a meaningful form
\begin{eqnarray}
 & & \hspace{-40pt}
\irn{3} \left[ 
\irn{3} \irn{3} \irn{3}  W_{\cE}(\ve, \vs | \vp, \vsp) 
\left( f' f'_{*} - f f_{*} \right)  d \vs \, d \vp \, d \vsp \right] 
\phi(\ve) \: d \ve \label{weakQ}
\\[7pt]
 & &  \hspace{-40pt}
\mbox{} = \frac{1}{2} \irn{3} \irn{3} \left[ \dfrac{}{}  
G(\phi; \ve , \vs) - \iiW(|\bV|) \left[ \phi(\ve) + \phi(\vs) \right] 
\right] \chi_{\cE}(\ve, \vs) \, f(\ve) \,  f(\vs) \: d \vs \, d \ve \p 
\label{eqGni}
\end{eqnarray}
where, as in the following, to simplify the notation, we omit to write the variables $t$
and $\bx$, explicitly.
In view of numerical calculations of \eq{eqGni}, we note that the function
\begin{equation}
G(\phi; \ve , \vs) - \iiW(|\bV|) \left[ \phi(\ve) + \phi(\vs) \right] ,
\label{kernwQ} 
\end{equation}
which depends on $(\ve, \vs)$ variables and also on the function $\phi$, 
plays the role of a kernel of the integral operator \eq{eqGni}.
So, \emph{if we are able to find a reasonable approximation of the function \eq{kernwQ}
for any fixed function $\phi$, then we can solve the six-fold integral
\eq{eqGni} instead of the twelve-fold integral \eq{weakQ}.}
There is a simple, but important, case where the function \eq{kernwQ} is known
explicitly. In fact, denoting by $\psi$ one of the collision invariants $1$, $\ve$,
$| \ve |^{2}$, we have
\begin{equation}
G(\psi; \ve , \vs) - \iiW(|\bV|) \left[ \psi(\ve) + \psi(\vs) \right] = 0
\quad \forall \, \ve , \vs \in \rea^{3} . \label{Gpsi} 
\end{equation}
%
%
\section{The operator G}
It is obvious that simple and explicit expressions of \eq{kernwQ} are unrealistic in the
whole space $\rea^{3} \times \rea^{3}$ and for generic $\phi$, apart from the case
\eq{Gpsi}. The scenario changes drastically, if the aim is to find
simple approximations of \eq{kernwQ} in small domains of the $(\ve, \vs)$ space and for a
finite set of $\phi$. Since the function $\iiW$ is usually given and does not require
further studies, we must consider only the operator $G$.

Let $D$ be a compact set of $\rea^{3}$ such that 
$\left\lbrace \ve \in \rea^{3} \: : |\ve|^{2}  \leq \cE \right\rbrace \subseteq D$.
We look for solutions of the Boltzmann equation vanishing for velocities outside the set
$D$.
We consider $N$ measurable sets $C_{\alpha}$ $(\alpha = 1, 2, ,.., N)$ such that
$$
C_{\alpha} \subseteq D \quad \forall \alpha \sv \quad
C_{\alpha} \cap C_{\beta} = \emptyset 
\quad \forall \, \alpha \neq \beta \sv \quad
\bigcup_{\alpha = 1}^{N} C_{\alpha} = D \p
$$
We denote by $\chi_{\alpha}$ the characteristic function on the set $C_{\alpha}$.

We are interesting to study $G(\psi \, \chi_{\gamma}; \ve , \vs)$.
We need to find a relationship similar to Eq.~\eq{kernwQ}.
Here we have
\begin{eqnarray*}
&&  \hspace{-20pt}
\sum_{\gamma = 1}^{N} G(\psi \, \chi_{\gamma}; \ve , \vs) =
\sum_{\gamma = 1}^{N}  \irn{3} \irn{3}  W(\ve, \vs | \vp, \vsp)
\left[ \psi(\vp) \, \chi_{\gamma}(\vp) + \psi(\vsp) \, \chi_{\gamma}(\vsp) \right] d \vp
\, d \vsp
\\
&&  \hspace{-20pt}
 =
\sum_{\gamma = 1}^{N} \left[  
\int_{C_{\gamma}} \! d \vp \irn{3} d \vsp \, W(\ve, \vs | \vp, \vsp) \, \psi(\vp) 
+ \irn{3} \! d \vp \int_{C_{\gamma}} d \vsp \, W(\ve, \vs | \vp, \vsp) \, \psi(\vsp)
\right]
\\
&&  \hspace{-20pt}
 =  
\int_{D} d \vp  \irn{3} d \vsp \, W(\ve, \vs | \vp, \vsp) \, \psi(\vp) 
+ \irn{3} d \vp \int_{D} d \vsp \, W(\ve, \vs | \vp, \vsp) \, \psi(\vsp) \p
\end{eqnarray*}
Moreover,
\begin{eqnarray*}
&&  \hspace{-20pt}
\chi_{\cE}(\ve, \vs) \, 
\sum_{\gamma = 1}^{N} G(\psi \, \chi_{\gamma}; \ve , \vs) 
\\
&&  \hspace{-20pt}
\mbox{} =  
\int_{D} d \vp  \irn{3} d \vsp \, W_{\cE}(\ve, \vs | \vp, \vsp) \, \psi(\vp) 
+ \irn{3} d \vp \int_{D} d \vsp \, W_{\cE}(\ve, \vs | \vp, \vsp) \, \psi(\vsp) 
\\
&&  \hspace{-20pt}
\mbox{} =  
\irn{3} d \vp \irn{3} d \vsp \, W_{\cE}(\ve, \vs | \vp, \vsp) \, \psi(\vp) 
+ \irn{3} d \vp \irn{3} d \vsp \, W_{\cE}(\ve, \vs | \vp, \vsp) \, \psi(\vsp) 
\\
&&  \hspace{-20pt}
\mbox{} =  \irn{3} \irn{3}  W_{\cE}(\ve, \vs | \vp, \vsp)
\left[ \psi(\vp) + \psi(\vsp) \right] d \vp \, d \vsp
\\
&&  \hspace{-20pt}
\mbox{} = \left[ \psi(\ve) + \psi(\vs) \right] 
\irn{3} \irn{3}  W_{\cE}(\ve, \vs | \vp, \vsp) \: d \vp \, d \vsp \p
\end{eqnarray*}
Thus
\begin{equation}
\chi_{\cE}(\ve, \vs) \, 
\sum_{\gamma = 1}^{N} G(\psi \, \chi_{\gamma}; \ve , \vs) =
 \left[ \psi(\ve) + \psi(\vs) \right] \chi_{\cE}(\ve, \vs) \, \iiW(|\bV|) \p
 \label{SGnu}
\end{equation}
\subsection{Some formulas}
Taking into account the results of Appendix A, it is easy to see that 
$ G(\phi; \ve, \vs)$ can be written, for a generic test function $\phi$, as follows
$$
\frac{1}{2} \irn{3}  K_{\bC}(\bC \cdot \bV, |\bV|) 
\left[ \phi \left( \oot (\ve + \vs) + \bC \right) +
\phi \left( \oot (\ve + \vs) - \bC \right) \right]
\delta \left( |\bC|^{2} - \mbox{$\frac{1}{4}$} |\bV|^{2} \right) d \bC \p
\label{GK}
$$
Hence $G(\psi \, \chi_{\gamma}; \ve , \vs)$ is the sum of the two integrals,
corresponding to the sign $+$ and $-$,
$$
\frac{1}{2}
\irn{3}  K_{\bC}(\bC \cdot \bV, |\bV|) \,
\psi \left( \oot (\ve + \vs) \pm  \bC \right) 
\chi_{\gamma} \left( \oot (\ve + \vs) \pm \bC \right)
\delta \left( |\bC|^{2} - \mbox{$\frac{1}{4}$} |\bV|^{2} \right) d \bC \p
$$
We now consider this operator for the five cases corresponding to the
five collision invariants $\psi$. We obtain the following expressions.
\\
{\sc Case: }$\psi = 1$
$$
\frac{1}{2}
\irn{3}  K_{\bC}(\bC \cdot \bV, |\bV|) \,
\chi_{\gamma} \left( \oot (\ve + \vs) \pm \bC \right)
\delta \left( |\bC|^{2} - \mbox{$\frac{1}{4}$} |\bV|^{2} \right) d \bC \p
$$
{\sc Case: }$\psi = \ve$
\begin{eqnarray*}
&&
\frac{1}{2}
\irn{3}  K_{\bC}(\bC \cdot \bV, |\bV|) \,
\left[ \oot (\ve + \vs) \pm  \bC \right]
\chi_{\gamma} \left( \oot (\ve + \vs) \pm \bC \right)
\delta \left( |\bC|^{2} - \mbox{$\frac{1}{4}$} |\bV|^{2} \right) d \bC
\\
&&
\mbox{} =
\frac{1}{4} \, (\ve + \vs)
\irn{3}  K_{\bC}(\bC \cdot \bV, |\bV|) \,
\chi_{\gamma} \left( \oot (\ve + \vs) \pm \bC \right)
\delta \left( |\bC|^{2} - \mbox{$\frac{1}{4}$} |\bV|^{2} \right) d \bC
\\
&&
\mbox{ } \pm
\frac{1}{2}
\irn{3}  K_{\bC}(\bC \cdot \bV, |\bV|) \, \bC \,
\chi_{\gamma} \left( \oot (\ve + \vs) \pm \bC \right)
\delta \left( |\bC|^{2} - \mbox{$\frac{1}{4}$} |\bV|^{2} \right) d \bC \p
\end{eqnarray*}
{\sc Case: }$\psi = |\ve|^{2}$
\\
Taking into account the Dirac distribution and replacing $|\bC|^{2}$ with 
$\frac{1}{4} |\bV|^{2}$, we have
$$
\left[ \oot \, (\ve + \vs) \pm  \bC \right]^{2} =
\oot \left[ |\ve|^{2} + |\vs|^{2} \right] \pm (\ve + \vs) \cdot \bC \p
$$
Then
\begin{eqnarray*}
&& \hspace{-15pt}
\frac{1}{2}
\irn{3}  K_{\bC}(\bC \cdot \bV, |\bV|) \,
\left[ \oot (\ve + \vs) \pm  \bC \right]^{2}
\chi_{\gamma} \left( \oot (\ve + \vs) \pm \bC \right)
\delta \left( |\bC|^{2} - \mbox{$\frac{1}{4}$} |\bV|^{2} \right) d \bC
\\
&& \hspace{-10pt}
\mbox{} =
\frac{1}{4} \left[ |\ve|^{2} + |\vs|^{2} \right]
\irn{3}  K_{\bC}(\bC \cdot \bV, |\bV|) \,
\chi_{\gamma} \left( \oot (\ve + \vs) \pm \bC \right)
\delta \left( |\bC|^{2} - \mbox{$\frac{1}{4}$} |\bV|^{2} \right) d \bC
\\
&& \hspace{-10pt}
\mbox{ } \pm
\oot \, (\ve + \vs) \cdot
\irn{3}  K_{\bC}(\bC \cdot \bV, |\bV|) \, \bC \,
\chi_{\gamma} \left( \oot (\ve + \vs) \pm \bC \right)
\delta \left( |\bC|^{2} - \mbox{$\frac{1}{4}$} |\bV|^{2} \right) d \bC \p
\end{eqnarray*}
In these cases, it is clear that the operator $G$ can be written using the two
couples of functions
\begin{eqnarray}
&& \hspace{-35pt}
A_{\gamma}^{\pm}(\ve, \vs) = 
\frac{1}{2} \, \irn{3}  K_{\bC}(\bC \cdot \bV, |\bV|) \,
\chi_{\gamma} \left( \oot (\ve + \vs) \pm \bC \right)
\delta \left( |\bC|^{2} - \mbox{$\frac{1}{4}$} |\bV|^{2} \right) d \bC
\label{Agamma}
\\
&& \hspace{-35pt}
\mathbf{B}_{\gamma}^{\pm}(\ve, \vs) =
\frac{1}{2} \, \irn{3}  K_{\bC}(\bC \cdot \bV, |\bV|) \, \bC \,
\chi_{\gamma} \left( \oot (\ve + \vs) \pm \bC \right)
\delta \left( |\bC|^{2} - \mbox{$\frac{1}{4}$} |\bV|^{2} \right) d \bC \p
\label{Bgamma}
\end{eqnarray}
Set
\begin{equation}
A_{\gamma}(\ve, \vs) = A_{\gamma}^{+}(\ve, \vs) + A_{\gamma}^{-}(\ve, \vs)
\mbox{ and }
\mathbf{B}_{\gamma}(\ve, \vs) =
\mathbf{B}_{\gamma}^{+}(\ve, \vs) - \mathbf{B}_{\gamma}^{-}(\ve, \vs) \, ,
\end{equation}
for  $\psi = 1$, $\ve$ and $|\ve|^{2}$, $G(\psi \, \chi_{\gamma}; \ve , \vs)$
writes 
\begin{eqnarray*}
 \left( \psi = 1 \right) & \rightarrow &
A_{\gamma}(\ve, \vs) 
\\[5pt]
\left( \psi = \ve \right)  & \rightarrow &
\frac{1}{2} \, (\ve + \vs) \, A_{\gamma}^{+}(\ve, \vs)
 + \mathbf{B}_{\gamma}(\ve, \vs) 
\\[5pt]
\left( \psi = |\ve|^{2} \right) & \rightarrow &
\frac{1}{2} \left[ |\ve|^{2} + |\vs|^{2} \right]
A_{\gamma}(\ve, \vs) 
+ \mathbf{B}_{\gamma}(\ve, \vs) \cdot (\ve + \vs) .
\end{eqnarray*}
\subsection{Approximation of G}
We look for an approximation of $G(\psi \, \chi_{\gamma}; \ve , \vs)$ for $\gamma = 1, 2,
.., N$ and any $\psi$.
In order to make clear the problem, we write part of Eq.~\eq{eqGni}
\begin{equation}
 \irn{3} \irn{3} 
G(\phi; \ve , \vs) \, \chi_{\cE}(\ve, \vs) \, f(\ve) \,  f(\vs) \: d \vs \, d \ve \p 
\label{Gff} 
\end{equation}
The objective is to find an approximation of the kernel
$G$ such that the error introduced by the new integral operator is small
for a reasonable set of distribution function $f$.

We note that $G(\psi \, \chi_{\gamma}; \ve , \vs)$ is given in
terms of the $A_{\gamma}(\ve, \vs)$ and $\mathbf{B}_{\gamma}(\ve, \vs)$; so, it
is sufficient to consider these functions.
Let $\alpha$, $\beta$ and $\gamma$ be three positive integers belonging to the interval
$[1, N]$. Suppose $M(\ve) : \rea^{3} \rightarrow \rea$ a positive function, which
represents a good candidate of the set of the functions $f$. We propose these
simple approximations
\begin{eqnarray}
& & \hspace{-25pt}
\chi_{\cE}(\ve, \vs) \, A_{\gamma}(\ve, \vs) \, M(\ve) \, M(\vs)
\approx
2 \,  \Phi_{\gamma \alpha \beta} \, \chi_{\cE}(\ve, \vs) \, 
\iiW(|\bV|) M(\ve) \, M(\vs)
\label{appA}
\\
& & \hspace{-25pt}
\chi_{\cE}(\ve, \vs) \, \mathbf{B}_{\gamma}(\ve, \vs) \, M(\ve) \, M(\vs)
\approx \bm{\Theta}_{\gamma \alpha \beta} \, \chi_{\cE}(\ve, \vs) \, |\bV| \,
\iiW(|\bV|) \, M(\ve) \, M(\vs) \label{appB}
\end{eqnarray}
$\forall \ve \in C_{\alpha}$ and $\vs \in C_{\beta}$.
In Eqs.~\eq{appA}-\eq{appB}, $\Phi_{\gamma \alpha \beta}$ and the array
$\bm{\Theta}_{\gamma \alpha \beta}$ are constant parameters to be determined. \\
The first step is to consider Eqs.~\eq{appA}-\eq{appB} only for a finite number of points
$\ve$ and $\vs$.
To this scope, we choose a finite set of well-distributed points in each cell
$C_{\alpha}$. We denote these points by $\ve_{\alpha i}$. \\
If
$
\chi_{\cE}(\ve_{\alpha i}, \ve_{\beta j}) \, |\ve_{\alpha i} - \ve_{\beta j}|
\, \iiW(|\ve_{\alpha i} - \ve_{\beta j}|) = 0
$
for every $i$ and $j$, then we define $ \Phi_{\gamma \alpha \beta} = 0$ and
$\bm{\Theta}_{\gamma \alpha \beta} = \bm{0}$. 
Otherwise, we use the standard least square method to find the parameters. 
Set $Y(\ve, \vs) = \chi_{\cE}(\ve, \vs) \, M(\ve) \, M(\vs)$, in our case the
problem is formulated as follows

\emph{find the minimum of the function of $\Phi_{\gamma \alpha \beta}$}
$$
\sum_{i,j} \left\lbrace
Y(\ve_{\alpha i}, \ve_{\beta j}) \left[ 
A_{\gamma}(\ve_{\alpha i}, \ve_{\beta j})
 -  2 \,  \Phi_{\gamma \alpha \beta} \, \iiW(|\ve_{\alpha i} - \ve_{\beta j}|)
\right] \right\rbrace^{2} ;
$$
\indent
\emph{and the minimum of the function of $\bm{\Theta}_{\gamma \alpha\beta}$}
$$
\sum_{i,j} \! \left\lbrace
Y(\ve_{\alpha i}, \ve_{\beta j}) \left[ 
\mathbf{B}_{\gamma}(\ve_{\alpha i}, \ve_{\beta j})
- \bm{\Theta}_{\gamma \alpha \beta}
\, |\ve_{\alpha i} - \ve_{\beta j}| \, \iiW(|\ve_{\alpha i} - \ve_{\beta j}|)
\right] \right\rbrace^{2} \! \! .
$$
The solution is obtained easily and it is given by
\begin{equation}
 \Phi_{\gamma \alpha \beta} =
\dfrac{ \dm \sum_{i,j} 
\left[ Y(\ve_{\alpha i}, \ve_{\beta j}) \right]^{2} 
A_{\gamma}(\ve_{\alpha i}, \ve_{\beta j}) \, \iiW(|\ve_{\alpha i} - \ve_{\beta j}|)
}{ \dm
2 \sum_{i,j} \left[ 
Y(\ve_{\alpha i}, \ve_{\beta j}) \, \iiW(|\ve_{\alpha i} - \ve_{\beta j}|)
\right]^{2} }  \label{sol_Phi} 
\end{equation}
and 
\begin{equation}
\bm{\Theta}_{\gamma \alpha \beta} =
\dfrac{ \dm \sum_{i,j} 
\left[ Y(\ve_{\alpha i}, \ve_{\beta j}) \right]^{2}
\mathbf{B}_{\gamma}(\ve_{\alpha i}, \ve_{\beta j}) \,
|\ve_{\alpha i} - \ve_{\beta j}| \, \iiW(|\ve_{\alpha i} - \ve_{\beta j}|)
}{ \dm
\sum_{i,j} \left[ 
Y(\ve_{\alpha i}, \ve_{\beta j}) \, |\ve_{\alpha i} - \ve_{\beta j}|
\, \iiW(|\ve_{\alpha i} - \ve_{\beta j}|)
\right]^{2} } . \label{sol_Theta} 
\end{equation}
It is simple matter to show that
\begin{equation}
\sum_{\gamma =1}^{N} \Phi_{\gamma \alpha \beta} = 1 \quad \mbox{ and } \quad
\sum_{\gamma =1}^{N} \bm{\Theta}_{\gamma \alpha \beta} = \bm{0} \p
\label{PT0} 
\end{equation}
The proof requires the use of the expressions of
$G(1 \chi_{\gamma}; \ve , \vs)$ and $G(\ve \, \chi_{\gamma}; \ve , \vs)$ in terms of
$A_{\gamma}(\ve, \vs)$ and $\mathbf{B}_{\gamma}(\ve, \vs)$,
and Eq.~\eq{SGnu}. 
It is also evident that the symmetry property 
$\Phi_{\gamma \alpha \beta} = \Phi_{\gamma \beta \alpha}$
and $ \bm{\Theta}_{\gamma \alpha \beta} =  \bm{\Theta}_{\gamma \beta \alpha}$
hold. \\
If we define
\begin{equation}
\mbox{} \hspace{-10pt}
R_{\gamma \alpha \beta}(\psi; \ve, \vs) =
|\bV| 
\left\lbrace \!
\begin{array}{ll}
0 & \mbox{for } \psi = 1 \\[5pt]
\bm{\Theta}_{\gamma \alpha \beta} & \mbox{for } \psi = \ve \\[5pt]
\bm{\Theta}_{\gamma \alpha \beta} \cdot  (\ve + \vs) & \mbox{for } \psi = |\ve|^{2} 
\end{array}
\right. 
(\ve \in C_{\alpha} \mbox{ and } \vs \in C_{\beta})
\end{equation}
then, for every $\gamma \in [1,N]$ and $\forall \ve \in C_{\alpha}$, $\vs \in C_{\beta}$,
we can  write the approximation of $G(\psi \, \chi_{\gamma}; \ve , \vs)$
in the following compact way
\begin{eqnarray}
&& \mbox{} \hspace{-60pt}
G(\psi \, \chi_{\gamma}; \ve , \vs) \, \chi_{\cE}(\ve, \vs) \approx
\chi_{\cE}(\ve, \vs) \, \iiW(|\bV|) \nonumber \\
&&
\mbox{} \hspace{50pt}
\left\lbrace
\Phi_{\gamma \alpha \beta} \left[ \psi(\ve) + \psi(\vs) \right] 
+ R_{\gamma \alpha \beta}(\psi; \ve, \vs) \right\rbrace .
\label{approG}
\end{eqnarray}
%
%
%
%
\section{Approximate weak form of the collisional operator}
Let us consider Eq.~\eq{eqGni}. We have
\\[7pt]
$ \dm
\irn{3} \left[ 
\irn{3} \irn{3} \irn{3}  W_{\cE}(\ve, \vs | \vp, \vsp) 
\left( f' f'_{*} - f f_{*} \right)  d \vs \, d \vp \, d \vsp \right] 
\psi(\ve) \, \chi_{\gamma}(\ve) \: d \ve
$
\\[7pt]
$ \dm
\mbox{} = \dfrac{1}{2} \irn{3} \irn{3} \left[ \dfrac{}{}
G(\psi \, \chi_{\gamma}; \ve , \vs) - \iiW(|\bV|) 
\left[ \psi(\ve) \, \chi_{\gamma}(\ve) + \psi(\vs) \, \chi_{\gamma}(\vs) \right] 
\right]
$
\\[7pt]
$ \dm
\mbox{} \hspace{50pt}
\chi_{\cE}(\ve, \vs) \, f(\ve) \,  f(\vs) d \vs \, d \ve
$
\\[7pt]
$ \dm
\mbox{} = \dfrac{1}{2} \sum_{\alpha, \beta} 
\int_{C_{\alpha}} \!  \!d \ve \int_{C_{\beta}} \! \! d \vs  \! \left[ 
\dfrac{}{} \! G(\psi \, \chi_{\gamma}; \ve , \vs) - \iiW(|\bV|) 
\left[ \psi(\ve) \, \chi_{\gamma}(\ve) + \psi(\vs) \, \chi_{\gamma}(\vs) \right] 
\right]  
$
\\[7pt]
$ \dm
\mbox{} \hspace{50pt}
\chi_{\cE}(\ve, \vs) \, f(\ve) \,  f(\vs)
$
\\[7pt]
$ \dm
\mbox{} = \dfrac{1}{2} \sum_{\alpha, \beta}
\int_{C_{\alpha}} d \ve \int_{C_{\beta}} d \vs \left[ \dfrac{}{}
G(\psi \, \chi_{\gamma}; \ve , \vs) - \iiW(|\bV|) 
\left[ \psi(\ve) \, \delta_{\gamma \alpha} + \psi(\vs) \, \delta_{\gamma \beta} \right]  
\right] 
$
\\[7pt]
$ \dm
\mbox{} \hspace{50pt}
\chi_{\cE}(\ve, \vs) \, f(\ve) \,  f(\vs) \qquad
$
(we now use the approximation \eq{approG} of $G$)
\\[7pt]
$ \dm
\mbox{} \approx \dfrac{1}{2} \sum_{\alpha, \beta}
\int_{C_{\alpha}} d \ve \int_{C_{\beta}} d \vs \: \iiW(|\bV|)
\left[\dfrac{}{} \Phi_{\gamma \alpha \beta} \left[ \psi(\ve) + \psi(\vs) \right] 
+ R_{\gamma \alpha \beta}(\psi; \ve, \vs) \right.
$
\\[7pt]
$ \dm \mbox{} \hspace{120pt} \left. \dfrac{}{}
- \left[ \psi(\ve) \, \delta_{\gamma \alpha} + \psi(\vs) \,
\delta_{\gamma \beta} \right] \right]
\chi_{\cE}(\ve, \vs) \, f(\ve) \,  f(\vs)
$
\\[7pt]
%
%
$ \dm
\mbox{} =
\dfrac{1}{2} \sum_{\alpha, \beta}
\left[ \Phi_{\gamma \alpha \beta} - \delta_{\gamma \alpha} \right]
\int_{C_{\alpha}} d \ve \int_{C_{\beta}} d \vs \: \iiW(|\bV|) \,
\psi(\ve) \, \chi_{\cE}(\ve, \vs) \, f(\ve) \,  f(\vs)
$
\\[7pt]
$ \dm
\mbox{} \quad +
\dfrac{1}{2} \sum_{\alpha, \beta}
\left[ \Phi_{\gamma \alpha \beta} - \delta_{\gamma \beta} \right]
\int_{C_{\alpha}} d \ve \int_{C_{\beta}} d \vs \: \iiW(|\bV|) \,
\psi(\vs) \, \chi_{\cE}(\ve, \vs) \, f(\ve) \,  f(\vs)
$
\\[7pt]
$ \dm
\mbox{} \quad +
\dfrac{1}{2} \sum_{\alpha, \beta}
\int_{C_{\alpha}} d \ve \int_{C_{\beta}} d \vs \: \iiW(|\bV|) \,
 R_{\gamma \alpha \beta}(\psi; \ve, \vs) \, \chi_{\cE}(\ve, \vs) \, f(\ve) \,  f(\vs)
$
\\[7pt]
%
%
$ \dm
\mbox{} =
\dfrac{1}{2} \sum_{\alpha, \beta}
\left[ \Phi_{\gamma \alpha \beta} - \delta_{\gamma \alpha} \right]
\int_{C_{\alpha}} d \ve \int_{C_{\beta}} d \vs \: \iiW(|\bV|) \,
\psi(\ve) \, \chi_{\cE}(\ve, \vs) \, f(\ve) \,  f(\vs)
$
\\[7pt]
$ \dm
\mbox{} \quad +
\dfrac{1}{2} \sum_{\alpha, \beta}
\left[ \Phi_{\gamma \alpha \beta} - \delta_{\gamma \alpha} \right]
\int_{C_{\beta}} d \ve \int_{C_{\alpha}} d \vs \: \iiW(|\bV|) \,
\psi(\vs) \, \chi_{\cE}(\ve, \vs) \, f(\ve) \,  f(\vs)
$
\\[7pt]
$ \dm
\mbox{} \quad +
\dfrac{1}{2} \sum_{\alpha, \beta}
\int_{C_{\alpha}} d \ve \int_{C_{\beta}} d \vs \: \iiW(|\bV|) \,
 R_{\gamma \alpha \beta}(\psi; \ve, \vs) \, \chi_{\cE}(\ve, \vs) \, f(\ve) \,  f(\vs) \p
$
\\[7pt]
Therefore we have
\begin{eqnarray}
& & \mbox{} \hspace{-10pt}
\irn{3} \left[ 
\irn{3} \irn{3} \irn{3}  W_{\cE}(\ve, \vs | \vp, \vsp) 
\left( f' f'_{*} - f f_{*} \right)  d \vs \, d \vp \, d \vsp \right] 
\psi(\ve) \, \chi_{\gamma}(\ve) \: d \ve
\nonumber \\[7pt]
& &
\mbox{} \approx \sum_{\alpha, \beta}
\left[ \Phi_{\gamma \alpha \beta} - \delta_{\gamma \alpha} \right]
\int_{C_{\alpha}} d \ve \int_{C_{\beta}} d \vs \: \iiW(|\bV|) \,
\psi(\ve) \, \chi_{\cE}(\ve, \vs) \, f(\ve) \,  f(\vs)
\nonumber \\[7pt]
& &
\mbox{ } +
\dfrac{1}{2} \sum_{\alpha, \beta}
\int_{C_{\alpha}} d \ve \int_{C_{\beta}} d \vs \: \iiW(|\bV|) \,
 R_{\gamma \alpha \beta}(\psi; \ve, \vs) \, \chi_{\cE}(\ve, \vs) \, f(\ve) \,  f(\vs)
\p \label{eqQappr}
\end{eqnarray}
This is the proposed model to approximate
$$
\irn{3} Q(f.f) \, \chi_{\gamma}(\ve) \, \psi(\ve) \: d \ve \p
$$
{\bf Remark 1.}
If we make the sum with respect to all values of $\gamma$, then the r.h.s of Eq.~\eq{eqQappr} vanishes
since $\psi$ is a collision invariant. Now, due to the properties \eq{PT0}, also the
l.h.s vanishes. Thus the modified collision operator guarantees the conservation of the
mass, momentum and energy as the true one.
\\
{\bf Remark 2.}
We have not introduced any constrain on the size and shape of the cell $C_{\alpha}$;
so, there is a great arbitrariness in the decomposition of the domain $D$.  
\\
{\bf Remark 3.}
We need the constant parameters $\Phi_{\gamma \alpha \beta}$ and 
$ \bm{\Theta}_{\gamma \alpha \beta}$, which depend only on the decomposition of $D$ and
the function $M(\ve)$; hence, they be evaluated once at beginning. 
The formulas \eq{sol_Phi} and \eq{sol_Theta} require to solve the integrals \eq{Agamma}
and \eq{Bgamma}. There is an high complexity, due to the presence both the characteristic
function $\chi_{\gamma}$ and the Dirac distribution. It can be eliminated by means of a
transformation of coordinates (from cartesian to spherical), by this changes the shape
of domain of integration. A very regular cell $C_{\gamma}$, having a simple geometry, is
usually transformed in a warped domain.
\emph{In our scheme we do not need accurate quadrature formulas}, because we have
Eqs.~\eq{PT0}.
These can used as a corrector. For instance, we can find the parameters 
$\Phi_{\gamma \alpha \beta}$, use \eq{PT0}$_{1}$ for checking the goodness and define the
new values by means of the simple formula
$$
\left( \Phi_{\gamma \alpha \beta} \right)_{new} =
\dfrac{\left( \Phi_{\gamma \alpha \beta} \right)_{old}}{
\dm
\sum_{\lambda=1}^{N} \left( \Phi_{\lambda \alpha \beta} \right)_{old}
} 
\qquad \forall \, \alpha , \beta, \gamma  \p
$$
We can use \eq{PT0}$_{2}$ to adjust the other parameters
$ \bm{\Theta}_{\gamma \alpha \beta}$.
%
%
%
%
\section{Approximate distribution functions}
In the framework of the discontinuous Galerkin method, a consistent approximation is given
using in each cell the same set of functions used as test functions. 
We denote by $\left\lbrace \psi_{j} : \: j=0,1..,4 \right\rbrace$ the ordered set
$\left\lbrace 1, \ve, |\ve|^{2} \right\rbrace$.
We now assume
\begin{equation}
f(t, \bx, \ve) \approx P(\ve) \sum_{\alpha=1}^{N} 
\left[ \mathbf{g}_{\alpha}(t, \bx) \cdot \bm{\eta}_{\alpha}(\ve) \right]
\chi_{\alpha}(\ve) \p
\label{approxf} 
\end{equation}
We have introduced a weight positive function $P(\ve)$, which we can choose taking into
account a reasonable or expected shape of the solution $f$.
The components of the five dimensional array $\bm{\eta}_{\alpha}(\ve)$ are functions,
denoted by $\eta_{\alpha, i}(\ve)$, which are linear combination of the collision
invariants and such that
\begin{equation}
\int_{C_{\alpha}} P(\ve) \, \eta_{\alpha, i}(\ve) \, \psi_{j}(\ve) \: d \ve = 
\delta_{i j} 
\end{equation}
for every $i$ and $j$ and for each cell $C_{\alpha}$.
The vector functions $\mathbf{g}_{\alpha}(t, \bx)$ $(\alpha = 1, 2, .., N)$ are the new
unknowns instead of the distribution function $f$.
It is immediate verify that
\begin{equation}
\int_{C_{\alpha}} f(t, \bx, \ve) \, \psi_{j}(\ve) \: d \ve \approx g_{\alpha, j}(t, \bx)
,
\end{equation}
where $g_{\alpha, j}(t, \bx)$ are the components of $g_{\alpha}(t, \bx)$.
Using the approximation \eq{approxf}, the integrals in the r.h.s. of Eq.~\eq{eqQappr}
become second degree polynomials in the new unknowns $\mathbf{g}_{\alpha}(t, \bx)$.
The numerical coefficients are constant and determined by solving the integrals
\begin{equation}
\dm \int_{C_{\alpha}} d \ve \int_{C_{\beta}} d \vs \:
\iiW(|\bV|) \, \psi(\ve) \,
\chi_{\cE}(\ve, \vs) \, P(\ve) \, P(\vs) \,
\eta_{\alpha, i}(\ve) \, \eta_{\beta, j}(\vs) 
\end{equation}
and
\begin{equation}
\dm \int_{C_{\alpha}} d \ve \int_{C_{\beta}} d \vs \:
\iiW(|\bV|) \, R_{\gamma \alpha \beta}(\psi; \ve, \vs) \,
\chi_{\cE}(\ve, \vs) \, P(\ve) \, P(\vs) \,
\eta_{\alpha, i}(\ve) \, \eta_{\beta, j}(\vs) \p
\end{equation}
Also these parameters can be find at the beginning because they depend only on the domain
decomposition and the weight function $P(\ve)$.
\section{Conclusions and future work}
We have proposed a numerical model for a deterministic treatment of the collision
operator in the framework of the discontinuous Galerkin method. The free streaming
term of the Boltzmann equation was studied \cite{BH}. 
A numerical application of the proposed technique requires the computation of a lot of
integrals. This work is required only at beginning. We made some very preliminary
tests, where we have used is a simple quadrature method. For instance, suppose that we
must integrate the product of two functions $p(z) \, q(z)$ where $z \in \rea^{m}$.
Let $R$ be the domain of integration and $R_{i}$ $(i = 1, 2, ... )$ a partition of $R$.
We denote by $z_{i}$ a suitable point in $R_{i}$. Then
$$
\int_{R} p(z) \, q(z) \: dz = \sum_{i} \int_{R_{i}} p(z) \, q(z) \: dz \approx
\sum_{i} p(z_{i}) \, \int_{R_{i}} q(z) \: dz \p
$$
This is useful, whether the integrals of the function $q$ are solved in $R_{i}$ 
analytically. The function $p$ is the ``bad'' part of the integral. In our case
the function $p$ contains characteristic functions and $\iiW(|\bV|)$ eventually, 
and $p$ is a polynomial in the velocity variables.

The validity of the numerical scheme will be considered in a future work.
Some very preliminary numerical experiments seem promising.
%
%
%
%
\section{Appendix A}
We derive a few formulas, which are present in the literature, both to introduce the
notation and to make the paper self-consistent.
Using the properties of Dirac distributions, we show how to reduce the order of
the integral  
\begin{equation}
\irn{3} \irn{3} F(\ve, \vs, \vp, \vsp) \,
\delta(\ve + \vs - \vp - \vsp) \,
\delta(|\ve|^{2} + |\vs|^{2} - |\vp|^{2} - |\vsp|^{2})
 \: d \vp \, d \vsp \p \label{i6F}
\end{equation}
Here $F$ is a generic function so that the following operations are allowed.
Introducing the change of variables
$\vp = \ve - \bm{A}$, $\vsp = \vs + \bm{B}$,
the integral \eq{i6F} becomes
$$
\irn{3} \irn{3} F(\ve, \vs, \ve - \bm{A}, \vs + \bm{B}) \,
\delta(\bm{A} - \bm{B}) \,
\delta(|\ve|^{2} + |\vs|^{2} - |\ve - \bm{A}|^{2} - |\vs + \bm{B}|^{2})
 \: d \bm{A} \, d \bm{B}
$$
and, after a simple integration,
$$
\frac{1}{2}
\irn{3} F(\ve, \vs, \ve - \bm{A}, \vs + \bm{A}) \,
\delta(|\bm{A}|^{2} - \bm{A} \cdot \bV) \: d \bm{A}  \p
$$
Here we have used the vector $\bV$, which is defined in \eq{nV}.
Now, performing the further change of variable
$\bC = \oot \bV  - \bm{A}$,
it is easy to verify that the integral \eq{i6F} is
\begin{equation}
\frac{1}{2}
\irn{3} F(\ve, \vs, \oot (\ve + \vs) + \bC, 
                  \oot (\ve + \vs) - \bC ) \,
\delta\left( |\bC|^{2} - \mbox{$\frac{1}{4}$} |\bV|^{2} \right) \: d \bC \p
\label{i3F} 
\end{equation}
One last reduction is available using the Dirac distribution in \eq{i3F}.
\subsection{The total cross section}
A simple application of \eq{i3F} allows to reduce the integral
$$
 \irn{3} \irn{3} W(\ve, \vs | \vp, \vsp) \: d \vp \, d \vsp
$$
to
\begin{equation}
\frac{1}{2} \irn{3} 
K \! \left( \bV \cdot \bm{n}_{\bC} , |\bV| \right) 
\delta \left( |\bC|^{2} - \mbox{$\frac{1}{4}$} |\bV|^{2} \right) 
 d \bC \, , \label{tcc}
\end{equation}
where
$$
\bm{n}_{\bC} = \dfrac{\ve - \oot (\ve + \vs) - \bC}{|\ve - \oot (\ve + \vs) - \bC|} \p
$$
With easy calculations and replacing $|\bC|^{2}$ with $\frac{1}{4} |\bV|^{2}$, we obtain
$$
\bV \cdot \bm{n}_{\bC} = \sqrt{\oot |\bV|^{2} - \bV \cdot \bC} \p
$$
Therefore we define
$$
 K_{\bC}(\bV \cdot \bC, |\bV|) = 
 K \! \left(\sqrt{\oot |\bV|^{2} - \bV \cdot \bC} \, , |\bV| \right) ,
$$
and we introduce a reference frame in $\rea^{3}$ and spherical coordinates such that
$$
\bC = r (\sin \varphi \, \cos \theta , \sin \varphi \, \sin \theta , \cos \varphi) 
\quad \mbox{and} \quad
\bC \cdot \bV = r \, |\bV| \cos \varphi \p
$$
Hence the integral \eq{tcc} reduces to
\begin{eqnarray*}
& &
\frac{1}{2} \int_{0}^{+ \infty} \hspace{-4pt}  dr 
\int_{0}^{\pi} \! d \varphi \int_{0}^{2 \pi} \! d \theta \,
K_{\bC}(r \, |\bV| \cos \varphi , |\bV|) \, 
\delta \left( r^{2} - \mbox{$\frac{1}{4}$} |\bV|^{2} \right) r^{2} \sin \varphi
\\
& & \mbox{ } = 
\frac{\pi}{4} |\bV| \int_{0}^{\pi}
K_{\bC}(\oot |\bV|^{2} \cos \varphi , |\bV|) 
\sin \varphi  \: d \varphi =
\frac{\pi}{4} |\bV| \int_{-1}^{1}
K_{\bC}(\oot |\bV|^{2} \mu , |\bV|) \: d \mu \p
\end{eqnarray*}

\begin{thebibliography}{99}
%
%
\bibitem{BH}
Lowell L. Baker and Nicolas G. Hadjiconstantinou, 
\emph{Variance-reduced Monte Carlo solutions of the Boltzmann
equation for low-speed gas flows: A discontinuous Galerkin formulation},
Int. J. Numer. Meth. Fluids, \textbf{58} (2008), 381--402.
%
\bibitem{carr03}
Jos\'{e} A. Carrillo, Irene M. Gamba, Armando Majorana and Chi-Wang Shu,
\emph{A WENO-solver for the transients of Boltzmann-Poisson system
for semiconductor devices. Performance and comparisons with Monte Carlo methods}, 
Journal of Computational Physics, \textbf{184} (2003), 498--525.
%
\bibitem{cgms06}
Jos\'{e} A. Carrillo, Irene M. Gamba, Armando Majorana and Chi-Wang Shu,
\emph{2D semiconductor device simulations by WENO-Boltzmann schemes:
efficiency, boundary conditions and comparison to Monte Carlo methods}, 
Journal of Computational Physics, \textbf{214} (2006), 55--80.
%
\bibitem{Cheng_08} 
Yingda Cheng, Irene M. Gamba, Armando Majorana and Chi-Wang Shu,
\emph{A discontinuous Galerkin solver for Boltzmann-Poisson systems
for semiconductor devices},
Computer Methods in Applied Mechanics and Engineering, \textbf{198} (2009),
3130--3150.
%
\bibitem{dgsurvey} 
B. Cockburn and C.-W. Shu, 
\emph{Runge-Kutta discontinuous Galerkin methods for convection-dominated
problems}, 
Journal of Scientific Computing, \textbf{16} (2001), 173--261.
%
\bibitem{RS} 
F. Rogier and J. Schneider,
\emph{A Direct Method for Solving the Boltzmann Equation},
Transport Theory Statist. Phys., \textbf{23} (1994), 313--338.
%
\bibitem{SOA}
Y. Sone,T. Ohwada and K. Aoki,
\emph{Temperature jump and Knudsen layer in a rarefied gas over a plane wall:
Numerical analysis of the linearized Boltzmann equation for hard‐sphere molecules},
Physics of Fluids A, \textbf{1} (1989), 363--370.
%
\bibitem{Aristov} 
V.V. Aristov,
``Direct methods for solving the Boltzmann equation and study of nonequilibrium
flows,''
Kluwer Academic Publishers, Boston, 2001.
%
\bibitem{C88} 
Carlo Cercignani,
``The Boltzmann Equation and its Applications''
Springer, New York, 1988.
%
\bibitem{C90} 
Carlo Cercignani,
``Mathematical Methods in Kinetic Theory''
\newblock  Plenum, New York, 1990.
%
\end{thebibliography}
\end{document}